\documentclass[amssymb]{revtex4}

\usepackage{graphicx}
\usepackage{epsfig}
\usepackage{dcolumn}
\usepackage{bm}

\def\gsim{\lower0.5ex\hbox{$\:\buildrel >\over\sim\:$}}
\def\lsim{\lower0.5ex\hbox{$\:\buildrel <\over\sim\:$}}

\begin{document}

\title{"Seesawing" away the hierarchy problem}

\author{Shaouly Bar-Shalom (speaker)\footnote{Talk given at the International Europhysics Conference on High Energy Physics, July 21st - 27th 2005, Lisboa, Portugal.}}
\email{shaouly@physics.technion.ac.il}
\affiliation{Physics Department, Technion-Institute of Technology, Haifa 32000, Israel}

\author{David Atwood}
\email{atwood@iastate.edu}
\affiliation{Department of Physics and Astronomy, Iowa State University, Ames,
IA 50011, USA}

\author{Amarjit Soni}
\email{soni@bnl.gov}
\affiliation{Theory Group, Brookhaven National Laboratory, Upton, NY 11973, USA}

\begin{abstract}
We describe a model for the scalar sector where all 
interactions occur
either at an ultra-high scale $\Lambda_U \sim 10^{16}-10^{19}$ GeV or 
at an intermediate scale $\Lambda_I=10^{9}-10^{11}$ GeV. 
The interaction of physics on these two scales 
results in an SU(2) Higgs condensate at the electroweak (EW) scale, 
$\Lambda_{EW}$, through a seesaw-like Higgs mechanism, 
$\Lambda_{EW} \sim \Lambda_I^2/\Lambda_U$, while the breaking of 
the SM SU(2)$\times$U(1) gauge symmetry 
occurs at the intermediate scale $\Lambda_I$.
The EW scale is, therefore, not fundamental but is
naturally generated in terms of ultra-high energy phenomena and so
the hierarchy problem is alleviated. We show that our 
``seesaw-Higgs'' model predicts the existence of 
sub-eV neutrino masses which are 
generated through
a ``two-step'' seesaw mechanism in terms of the same 
two ultra-high scales: 
$m_\nu \sim \Lambda_I^4/\Lambda_U^3 \sim \Lambda_{EW}^2/\Lambda_U $.
We also show that our seesaw Higgs model
can be naturally embedded in theories with {\it tiny} extra dimensions
of size $R \sim \Lambda_U^{-1} \sim 10^{-16}$ fm, 
where the seesaw induced EW scale 
arises from a violation of a symmetry at a distant brane if 
there are 7 {\it tiny} extra dimensions.
\end{abstract}

\maketitle

\section{Introduction}
A long standing problem in modern particle physics is the 
apparent enormous hierarchies of energy/mass scales observed 
in nature. Disregarding the ``small'' hierarchies in the masses of the 
known charged matter particles, there seems to be two much larger hierarchies:
the first is the hierarchy between the fundamental grand unified scale 
$\Lambda_U \sim {\cal O}(10^{16})$ GeV [or Planck scale 
$\Lambda_U \sim {\cal O}(10^{19})$ GeV], and the EW scale, 
$\Lambda_{EW} \sim {\cal O}(100)$ GeV, and the second is the   
hierarchy between the EW scale and the neutrino mass scale 
$m_\nu \sim {\cal O}(10^{-2})$ eV. 
This apparent hierarchical structure of scales
have fueled a lot of activity in the past 30 years in the search for new 
physics beyond the Standard Model (SM).

The $\Lambda_U - \Lambda_{EW}$ hierarchy, 
when viewed within the SM framework, is usually referred to as the 
gauge Hierarchy Problem (HP) of the SM, which is intimately related to 
the SM Higgs sector responsible for the generation of the 
EWSB scale, $v_{EW} \sim \Lambda_{EW}$, through the SM Higgs mechanism. 
The HP of the SM raises a technical difficulty known as the naturalness 
(or fine tuning) problem, i.e., 
there is a problem of stabilizing the ${\cal O}(\Lambda_{EW})$ 
mass scale of the Higgs against radiative corrections without 
an extreme fine tuning (to one part in $\Lambda_{EW}^2/\Lambda_U^2$).
It should, however, be emphasized that this fine-tuning problem of the SM
may be just a technical difficulty which reflects our ignorance in explaining 
the simultaneous presence of the two disparate scales 
$\Lambda_U$ and $\Lambda_{EW}$, and {\it may have nothing to do} 
with the 
more fundamental question of the origin of scales which we will 
address in this work: 
{\it why do we observe in nature such large hierarchies between the 
fundamental GUT or Planck scale $\Lambda_U$, the EW scale $\Lambda_{EW}$ 
and the neutrino mass-scale $m_\nu$}?

In this letter we propose a simple model \cite{paper}, 
where the only fundamental scale is the GUT or Planck 
scale $\Lambda_U$, while the EW and neutrino mass scales both arise 
due to interactions between
this fundamental scale $\Lambda_U$ and a new intermediate ultra-high scale
$\Lambda_I \sim 10^{9}-10^{11}$ GeV, i.e., 
$\Lambda_{EW}<<\Lambda_I<<\Lambda_U$. The intermediate scale   
is viewed as the
scale of breaking of the unification group which underlies the physics at 
the scale $\Lambda_U$ (see e.g., \cite{mohapatra}).   
Our model then naturally accounts for the existence of both the EW and sub-eV 
neutrino mass scales by means of a ``two-step'' seesaw between the two 
ultra-high mass scales $\Lambda_U$ and $\Lambda_I$: 
the first $\Lambda_U-\Lambda_I$ seesaw generates the
EW scale $\Lambda_{EW} \sim \Lambda_I^2/\Lambda_U$ and then a 
second $\Lambda_U-\Lambda_{EW}$ seesaw gives 
rise to the sub-eV neutrino mass scale 
$m_\nu \sim \Lambda_{EW}^2/\Lambda_U \sim \Lambda_I^4/\Lambda_U^3$.
Our model does not address the fine tuning problem - we assume that some 
higher symmetry at the fundamental scale $\Lambda_U$ is responsible 
for protecting the EW Higgs mass scale.

\section{The seesaw-Higgs model}

Let us schematically define the Lagrangian of our seesaw-Higgs model as 
follows:
\begin{eqnarray}
{\cal L}={\cal L}_{SM}(f,G)+ {\cal L}_{Y}(\Phi,f) + 
{\cal L}_S(\Phi,\varphi,\chi) + 
{\cal L}_\nu(\Phi,\varphi,\chi,\nu_L,\nu_R) 
\label{totlag} ~,
\end{eqnarray}   
\noindent where $\Phi$ is an SU(2) scalar doublet and $\varphi$, 
$\chi$ are ``sterile'' SU(2)-singlets that do not interact with the SM 
particles. Also, ${\cal L}_{SM}(f,G)$ contains the usual SM's fermions and 
gauge-bosons kinetic terms, ${\cal L}_{Y}(\Phi,f)$ contains the SM-like 
Yukawa interactions and
\begin{eqnarray}
{\cal L}_\nu(\Phi,\varphi,\chi,\nu_L,\nu_R) &=& - Y_D \ell_L \Phi \nu_R + 
Y_M \varphi \bar{\nu_R^c} \nu_R + Y^\prime_M \chi \bar{\nu_R^c} \nu_R+ h.c. 
~, \\
{\cal L}_S(\Phi,\varphi,\chi) &=& |D_\mu \Phi|^2 + |\partial_\mu \varphi|^2 +
|\partial_\mu \chi|^2 - V ~,
\end{eqnarray}
\noindent with 
\begin{eqnarray}
V = \lambda_1 \left( |\Phi|^2 - |\chi|^2 \right)^2 +
\lambda_2 \left( |\varphi|^2 - \Lambda_U^2 \right)^2 
+  ~ \lambda_3 \left( {\rm Re}(\varphi^\dagger \chi) - \Lambda_I^2 
\right)^2 + ~ \lambda_4 \left({\rm Im}(\varphi^\dagger \chi) - \Lambda_I^2 
\right)^2 \label{hpotlittle} ~,   
\end{eqnarray}
\noindent where all
$\lambda_i$ are positive real constants, naturally of ${\cal O}(1)$.
Note that the above total Lagrangian conserves lepton number $L$ 
if we assign lepton number 2 to both singlets $\varphi$ and $\chi$, 
i.e., if $L_\varphi=L_\chi=2$.

\section{The seesaw-Higgs mechanism and the Electroweak scale}

The seesaw-Higgs potential in Eq.~\ref{hpotlittle} gives rise to
the desired seesaw-condensate of $\Phi$.
In particular, the minimization of $V$ 
which only contains terms at energy scales $\Lambda_U$ and $\Lambda_I$ 
leads to:
\begin{eqnarray}
<\varphi>  &=& \Lambda_U ~,\nonumber\\ 
<\Phi> &=& <\chi> = \frac{\Lambda_I^2}{\Lambda_U} 
\equiv v_{EW} \sim \Lambda_{EW}
\label{seesawvev1} ~,
\end{eqnarray}      
\noindent where $<\Phi>=v_{EW}= \Lambda_I^2/\Lambda_U$ is the 
condensate required for EWSB, when 
the fundamental scale $\Lambda_U$ is taken to be 
around the GUT scale, $\Lambda_U \sim {\cal O}(10^{16})$ GeV, 
and the intermediate scale is $\Lambda_U \sim {\cal O}(10^{9})$ GeV, 
or when $\Lambda_U \sim {\cal O}(10^{19})$ GeV (the Planck scale) and 
$\Lambda_U \sim {\cal O}(10^{10.5})$ GeV.  

After EWSB 
we are left with 5 physical neutral scalars: 
$H$ which is 
a SM-like light Higgs with a mass $M_H \sim 2\sqrt{\lambda_1} v_{EW}$, 
3 superheavy 
neutral states $S_1,~S_2,~A_1$ with masses 
$M_{S_1} \sim \sqrt{\lambda_3} \Lambda_U$,
$M_{S_2} \sim 2 \sqrt{\lambda_2} \Lambda_U$, 
$M_{A_1} \sim \sqrt{\lambda_4} \Lambda_U$
and 1 massless axial state $A_M$ which 
is the ``Majoron'' \cite{plb98p265} associated with the
spontaneous breakdown of Lepton number (by the condensate of the 
two singlets, see next section).

\section{A two-step seesaw and the neutrino mass scale}

When the singlet $\varphi$ forms its 
condensate, $<\varphi> = \Lambda_U$, the second term in 
${\cal L}_\nu(\Phi,\varphi,\chi,\nu_L,\nu_R)$ will lead 
to a right-handed Majorana mass which will naturally be of that 
order: $m_\nu^M=Y_M \Lambda_U$.$^{[1]}$\footnotetext[1]{Note 
that, since $\chi$ forms a condensate of ${\cal O}(\Lambda_{EW})$, its 
contribution to the Majorana neutrino mass term will be negligible compared 
to that of $\varphi$ which forms the condensate of ${\cal O}(\Lambda_U)$.} 
The SU(2) condensate $<\Phi>=\Lambda_I^2/\Lambda_U \sim \Lambda_{EW}$ 
will generate a Dirac mass for the neutrinos 
of size $m_\nu^D \sim Y_D \Lambda_{EW}$ through the first term 
in ${\cal L}_\nu(\Phi,\varphi,\chi,\nu_L,\nu_R)$.  
Then, the neutrino mass matrix acquires the classic seesaw structure 
which, upon diagonalization, yields 
two physical Majorana neutrino states: 
a superheavy state $\nu_h$ with a mass 
$m_{\nu_h} \sim Y_M \Lambda_U$ and a superlight 
state $\nu_\ell$ with a mass:
\begin{eqnarray}
m_{\nu_\ell}=\frac{(m_\nu^D)^2}{m_\nu^M}
=\frac{Y_D^2}{Y_M} 
\frac{\Lambda_I^4}{\Lambda_U^3}
\sim\frac{Y_D^2}{Y_M} 
\frac{\Lambda_{EW}^2}{\Lambda_U}
\label{nuscale} ~.
\end{eqnarray}
\noindent The neutrino mass scale is, therefore, 
subject to a two-step seesaw 
mechanism,   
the first (in the scalar sector) generates the Dirac neutrino mass 
$m_\nu^D \sim Y_D \Lambda_{EW}$, which 
then enters in the off diagonal neutrino mass matrix to give the classic 
``seesawed'' Majorana mass in (\ref{nuscale}) by a second $m_\nu^M - m_\nu^D$ 
seesaw in the neutrino mass matrix.   
The presence of this extremely small scale, 
$m_{\nu_\ell} \sim {\cal O}(\Lambda_{EW}^2/\Lambda_U)$, well 
below the EW
scale, is therefore naturally explained in terms of the two ultra-high
scales $\Lambda_U$ and $\Lambda_I$.
For example, if $\Lambda_U \sim {\cal O}(10^{16})$ GeV
and $\Lambda_I \sim {\cal O}(10^{9})$ (which gives 
$\Lambda_{EW} \sim {\cal O}(100)$ GeV) we obtain for 
$Y_D \sim Y_M \sim {\cal O}(1)$:  
$m_{\nu_\ell} \sim {\cal O}(10^{-3})$ eV, 
roughly in accord with current mixing data.
A value of $\Lambda_U$ at the Planck scale could still be consistent 
with the double-seesaw sub-eV neutrino masses, if 
$\Lambda_I={\cal O}(10^{10.5})$ GeV (again giving 
$\Lambda_{EW} \sim {\cal O}(100)$ GeV) when 
$Y_D^2 / Y_M \sim {\cal O}(10^{3})$ GeV.
This may happen if e.g., 
the heavy Majorana mass term is of the order of the 
intermediate scale $\Lambda_I$, 
and the Dirac mass term is of 
${\cal O}(100)$ MeV  
(consistent with most light leptons and down quark masses).

\section{The seesaw-Higgs model from tiny extra dimensions}

If there are extra compact spatial dimensions (ECSD) 
which are populated with multiple 3-branes, 
then, as was shown in \cite{multibranes}, 
the violation of flavor symmetries on these distant branes can be
carried out to our brane by "messenger" scalar fields that can propagate 
freely in the bulk between the branes. In particular,
the profile of these messenger fields at all points on our wall (i.e., 
on the interference between the bulk and our brane)  
``shines'' the flavor violation which 
appears as a boundary condition on our $3$-brane.  

In our case, 
this ``shining'' mechanism can be utilized
to generate the seesaw-Higgs potential through the 
``shined'' value of the condensate of a messenger 
field $\eta$ on our wall \cite{paper}:
\begin{eqnarray}
<\eta> \sim \frac{\Gamma(\frac{n-2}{2})}{4 \pi^{\frac{n}{2}}}
\frac{M_\star}{(M_\star R)^{n-2}} \label{shine} ~,  
\end{eqnarray}
\noindent where $M_\star \sim \Lambda_U$ is the fundamental $4+n$
mass scale and $R$ is the size of the ECSD.   
In particular, an interaction term on our wall of the form: 
\begin{eqnarray}
S_{us} = \int_{us} d^4 x ~
\eta(x,y^i=0) \eta(x,y^i=0) \varphi^\dagger(x) \chi(x) +h.c. 
\label{sus}~.
\end{eqnarray}  
\noindent will yield the term $\Lambda_I^2 \varphi^\dagger \chi$ 
of the seesaw-Higgs potential in (\ref{hpotlittle}), if $<\eta> = \Lambda_I$.
Thus, using (\ref{shine}) with $M_\star \sim 10^{16}$ GeV, 
we find that the desired 
intermediate scale (i.e., $<\eta> = \Lambda_I \sim 10^9$ GeV 
in order to get the seesaw-induced EW scale) is obtained 
if there are $n=7$ {\it tiny extra transverse dimensions} of size 
$R \sim M_\star^{-1} \sim 10^{-16}$ fm.



\end{document}